\documentclass[aps,prd,reprint,twocolumn,superscriptaddress,longbibliography,nofootinbib,floatfix]{revtex4-1}
\usepackage{epsfig}
\usepackage{amsmath,amssymb,amsfonts}
\usepackage{hyperref}
\usepackage{mathrsfs}
\usepackage{bbm}
\usepackage{slashed}
\usepackage{graphicx}
\usepackage{verbatim}

\usepackage{stmaryrd}

\usepackage[usenames]{color}

\newcommand{\ddelta}{T}

\newcommand{\be}{\begin{equation}}
\newcommand{\ee}{\end{equation}}
\newcommand{\bw}{\begin{widetext}}
\newcommand{\ew}{\end{widetext}}
\newcommand{\bi}{\begin{itemize}}
\newcommand{\ei}{\end{itemize}}
\newcommand{\bea}{\begin{eqnarray}}
\newcommand{\eea}{\end{eqnarray}}
\newcommand{\bra}[1]{\langle\,#1\,|}          
\newcommand{\ket}[1]{|\,#1\,\rangle}          
\newcommand{\ud}{\mathrm{d}}

\newcommand{\LCp}{{\scriptscriptstyle +}}

\newcommand{\LCperp}{{\scriptscriptstyle \perp}}

\usepackage[T1]{fontenc} \usepackage[latin1]{inputenc}

\newcommand{\tint}[1]{\int\!\ud{\sf #1}}

\begin{document}

\title{Photon polarisation in light-by-light scattering: finite size effects}

\author{Victor Dinu}
\email[]{dinu@barutu.fizica.unibuc.ro}
\affiliation{Department of Physics, University of Bucharest, P. O. Box MG-11, M\u agurele 077125, Romania}

\author{Tom Heinzl}
\email[]{theinzl@plymouth.ac.uk}
\affiliation{School of Computing and Mathematics, Plymouth University, Plymouth PL4 8AA, UK}

\author{Anton Ilderton}
\email[]{anton.ilderton@chalmers.se}
\affiliation{Department of Applied Physics, Chalmers University of Technology, SE-41296 Gothenburg, Sweden}

\author{Mattias Marklund}
\email[]{mattias.marklund@chalmers.se}
\affiliation{Department of Applied Physics, Chalmers University of Technology, SE-41296 Gothenburg, Sweden}

\author{Greger Torgrimsson}
\email[]{greger.torgrimsson@chalmers.se}
\affiliation{Department of Applied Physics, Chalmers University of Technology, SE-41296 Gothenburg, Sweden}

\begin{abstract}
We derive a simple expression for the photon helicity and polarisation-flip probabilities in arbitrary background fields, in the low energy regime. Taking the background to model a focused laser beam, we study the impact of pulse shape and collision geometry on the probabilities and on ellipticity signals of vacuum birefringence. We find that models which do not account for pulse duration can overestimate all signals in near head-on collisions by up to an order of magnitude. Taking pulse duration into account, the flip probability becomes relatively insensitive to both angular incidence and the fine details of the pulse structure.

\end{abstract}

\maketitle

\section{Introduction}

It has been known since the early days of quantum electrodynamics (QED) that the appearance of virtual pairs leads to nonlinearities, due to the possibility of light-by-light scattering~\cite{Halpern:1934,Euler:1935zz,Heisenberg:1935qt}. These nonlinearities can manifest as effects akin to those in nonlinear optics~\cite{Marklund:2006my}: for example, a macroscopic, classical light source of sufficiently high intensity can alter the polarisation state of probe photons, leading to `vacuum' birefringence~\cite{Toll:1952}. (Note: `vacuum' here highlights only the absence of matter.)

In~\cite{Heinzl:2006xc} a proposal was made to demonstrate these nonlinearities. Because the birefringence effect increases with (target) field strength and probe frequency, it was suggested to use an intense, high-power laser as a target (implying a gain in field strength of many orders of magnitude compared to experiments using magnets~\cite{Zavattini:2012zs,Cadene:2013bva}) and an X-ray free electron laser (XFEL) as a probe. This scenario will be realised with the HIBEF facility employing the European XFEL at DESY~\cite{HIBEF}. Even higher intensities and probe energies will be achieved after the completion of the Extreme Light Infrastructure nuclear physics pillar (ELI-NP)~\cite{ELI}. The search for vacuum birefringence at HIBEF has been selected as its flagship experiment. It thus seems timely to extend the results of \cite{Heinzl:2006xc} by considering more realistic background field distributions modelling focused, pulsed lasers. This will also provide some theoretical underpinning for a detailed experimental feasibility study that is currently under way~\cite{HP}. 

To further motivate our investigation, recall that a beam of light, wavelength $\lambda'$, probing a birefringent medium acquires an ellipticity $\delta$ in its polarisation. In optics, $\delta$ can be expressed in terms of the refractive indices $\{n_\LCperp,n_\parallel\}$ of the medium, and the distance $d$ travelled in the medium by the probe as $\delta =  \pi d(n_\perp - n_\parallel)/\lambda'.$ This expression also holds for vacuum birefringence, under the assumptions that 1) this is induced by a homogenous, constant field, and 2) the probe is a plane wave. However, the targets and probes in upcoming experiments are lasers, and all fields will be focused, pulsed, and varying in space and time. Because the ellipticity is small and will be challenging to measure, a comprehensive discussion of potential experiments requires more careful modelling of the target and probe (as well as an analysis of background noise, losses in polarisers and lenses, and so on \cite{HP}). As a step toward this goal, the first purpose of this paper is to provide some simple but accurate formulae describing the impact of pulse shape and duration.

As stated above, and emphasised in~\cite{Dinu:2013gaa}, vacuum birefringence is a manifestation of photon-photon scattering. Hence, a measurement of the former would represent the first observation of the latter in a set-up with \emph{all} photons involved being real (unlike, say, in Delbr\"uck scattering~\cite{Jarlskog:1974tx,Schumacher:1975kv}). For real photon-photon scattering there are currently only upper bounds on the cross section~\cite{Bernard:2010dx}. The QED scattering processes underlying vacuum birefringence are therefore of interest~\cite{Davila:2013wba,Mo}, and it is natural to take an {\it S}-matrix approach to this topic. This was the approach taken in~\cite{Dinu:2013gaa}, in which we showed that the most relevant process is that in which probe photons flip between orthogonal helicity or polarisation states when passing through the target field. For the analytically solvable case of plane wave targets and probes, we obtained the helicity flip probability and the resulting probe ellipticity for arbitrary energies and intensities. The second purpose of this paper is to extend those results to cover backgrounds describing focused laser pulses in the relevant parameter regime; we study here the flip probability for photons probing intense, focused laser fields.

The paper is organised as follows. In Section~\ref{SECT:WORLDLINE} we derive the flip probability from the QED {\it S}-matrix; it takes the form of a simple integral over the worldline of a massless particle. In Sections~\ref{SECT:BEAMS} and~\ref{VD} we investigate the impact of field and collision geometries on the flip probability, and describe the implications for detecting ellipticity signals of vacuum birefringence. We conclude in Section~\ref{SECT:CONCLUSIONS}.

\section{A worldline integral for helicity and polarisation flip}\label{SECT:WORLDLINE}
We begin with the probability for a photon, momentum $l_\mu$, to flip helicity state $\epsilon_\mu\to\epsilon'_\mu$ when passing through a given background field  $f_{\mu\nu}$.  For the particular case of a plane wave background depending on $nx$ with $n^2=0$, a direct QED calculation of the amplitude, to 1-loop order and exact in all other parameters, was given in~\cite{Dinu:2013gaa}.  In the low energy regime relevant to laser-based experiments, the scattering amplitude reduces to
\be\begin{split}\label{worldline}
	\ddelta &= \frac{\alpha}{30\pi}\frac{1}{E_S^2}\int\!\frac{\ud (nx)}{nl} \big(lf(x)\bar{\epsilon}'\big) \big(lf(x)\epsilon\big) \;.
\end{split}
\ee
Our first task is to generalise this to arbitrary $f_{\mu\nu}$. To do so, recall first that a massless particle, momentum $l_\mu$, follows a null geodesic in spacetime. In Minkowski space, any `lightfront time' $nx$ provides a suitable affine parameterisation of the geodesic~\cite{ZMP}, the explicit form of which is
\be\label{bana}
	x_l^\mu(nx)= x^\mu(0) + \frac{l^\mu}{nl}\, nx \implies \ud x_l^\mu = \frac{l^\mu}{nl}\, \ud (nx) \;.
\ee
Using this measure in (\ref{worldline}) (extracting a factor of $l_\mu$ from the integrand turns the measure into $\ud x_l^\mu$ more explicitly) and taking $f_{\mu\nu}$ to be arbitrary and evaluated on the path (\ref{bana}), gives us a candidate expression for the flip amplitude in arbitrary backgrounds. Remarkably, this worldline integral is the correct expression, as we confirm below. Here we describe the physics it contains, beginning by writing it in a more revealing form.

The integral (\ref{worldline}) also gives the amplitude for a photon to flip between {\it any} two orthogonal polarisation states with $\epsilon'\epsilon=0$ (and $\epsilon^2={\epsilon'}^2=-1$) not only helicity states~\cite{Dinu:2013gaa}. We are interested in linearly polarised probes for birefringence, so we take $\{\epsilon,\epsilon'\}$ to be real; $\ddelta$ is then real. Recall that it is possible to choose polarisation vectors which are orthogonal to both $l_\mu$ and a second lightlike $n_\mu$~\cite{Brodsky:1997de,Heinzl:2000ht}, and so the vectors in play form a tetrad,
\be\label{tetrad}
	g_{\mu\nu} = \frac{n_\mu l_\nu + l_\mu n_\nu}{nl} - \epsilon_\mu \epsilon_\nu - \epsilon'_\mu \epsilon'_\nu \;.
\ee
Using this and the background energy-momentum tensor $\Theta_{\mu\nu} = f^2_{\mu\nu} -g_{\mu\nu}  \tfrac{1}{4}\text{tr} f^2$,  we can rewrite (\ref{worldline}) as 
\be\label{worldline2}
	\ddelta = \frac{\alpha}{60\pi}\frac{1}{E_S^2}\int\!\frac{\ud(nx)}{nl}\ \bigg[ l\Theta(x_l)l - (lf(x_l)\tilde\epsilon)^2\bigg] \;,
\ee
with $\tilde\epsilon:=\epsilon-\epsilon'$. The first term in the integrand of~(\ref{worldline2}) is the null projection of the energy-momentum tensor, as seen locally by the probe. Its appearance is to be expected~\cite{Shore:2007um} and maximising it maximises the probability (since the amplitude is now real). The second term in (\ref{worldline2}) is negative, and therefore reduces the amplitude, but can be made to vanish for appropriate choices of collision and polarisation geometries. The combination $lf\epsilon$ which appears is typical of polarisation transport~\cite{Bargmann:1959gz}.

As it should be, (\ref{worldline2}) is reparameterisation invariant (most easily seen by extracting a factor of $l_\mu$ and writing the meaure as $\ud x^\mu_l$) and gauge invariant (shifting the polarisation vectors by $l_\mu$ does not affect the amplitude). The integral is taken not over time, or position, but over the worldline of a massless particle, and so is fully relativistic. The form of (\ref{worldline2}) is similar to that of the eikonal found in high-energy scattering at small momentum transfer, which is also given by an integral over a classical particle trajectory, see~\cite[\S 9.1.1]{Itzykson:1980rh} or \cite[\S 9.6]{ZJ}. Here the presence of a worldline (rather than spacetime) integral encodes the possibility that $\ddelta$ vanishes when the photon misses a compactly supported background. The photon momentum $l_\mu$ is on-shell, underlining that we have abandoned the effective approach in favour of a `microscopic' approach, and constant, because only the forward-scattering flip amplitude is relevant in the considered regime. This will be explained below, when we derive (\ref{worldline2}) from the $S$-matrix. The reader primarily interested in phenomenology may proceed directly to Sect.~\ref{SECT:BEAMS}, where we evaluate the worldline integral.
  
\subsection{Derivation from low-energy scattering in QED}\label{SECT:HE}
To derive (\ref{worldline}) from QED, we would first write down the one-loop {\it S}-matrix element for helicity flip in a background field, and integrate out the fermions (giving the polarisation tensor, see e.g.~\cite{Narozhny:1968, Becker-Mitter, Baier:1975ff, Meuren:2013oya,Karbstein:2013ufa} and references therein). Though this cannot be done analytically for arbitrary backgrounds and arbitrary probe frequencies, we note that (\ref{worldline}) is a low-energy approximation, of the same order as if we had treated the background perturbatively~\cite{Euler:1935zz,Heisenberg:1935qt,DiPiazza:2013iwa}, and that there are no derivatives on the field. Therefore, given the approximations involved, it is simplest to begin with the low-energy Heisenberg-Euler effective action~\cite{Euler:1935zz,Heisenberg:1935qt}, see also~\cite[\S5.1]{Dittrich:2000zu},
\be\label{LHE}
	\mathcal{L}_{HE}= \frac{1}{4} \text{tr}\, F^2 + \frac{2\alpha^2}{45m^4}\bigg[ \frac{7}{4} \text{tr}\, F^4 - \frac{5}{8}(\text{tr}\, F^2)^2\bigg] \;.
\ee
We have written, for example, $\text{tr}\, F^2 = {F^{\mu}}_\nu {F^{\nu}}_\mu = - F^{\mu\nu}F_{\mu\nu}$ and we have used the formula~\cite{Davila:2013wba}
\be
	(\text{tr}\, F\tilde{F})^2 = 4 \text{tr}\, F^4 - 2 (\text{tr}\, F^2)^2 \;,
\ee
to remove the dual tensor. The background $f$ is introduced by replacing $eF \to eF + ef$, and retaining only terms which are quadratic in both $F$ and $f$ (other terms do not contribute here).

We will calculate the probability for an incoming photon to scatter, momentum $p_\mu\to l'_\mu$, and flip polarisation, $\epsilon\to \epsilon'$ with $\epsilon'\epsilon=0$. The relevant {\it S}-matrix element is obtained as usual from LSZ reduction of the correlation functions generated using the Lagrangian~(\ref{LHE}). It takes the form typical of scattering from an external, classical potential, in this case $f^2_{\mu\nu}$,
\be\label{SFI}
	\bra{l',\epsilon'}S\ket{p,\epsilon}=\int\!\ud^4x \; e^{i(l'-p)x} \llbracket F_\text{in}F_\text{out}f^2(x) \rrbracket \;,
\ee
in which $x$ is the vertex position, $F_\text{in}^{\mu\nu}=p^{[\mu}\epsilon^{\nu]}$, $F_\text{out}^{\mu\nu}=l'^{[\mu}\epsilon'^{\nu]}$ and $\llbracket F_\text{in}F_\text{out}f^2 \rrbracket$ is shorthand for the (several) trace terms in $\mathcal{L}[F_\text{in}+F_\text{out}+f]$ which are quadratic in $f$ and linear in $F_\text{in}$ and $F_\text{out}$.

The states in (\ref{SFI}) are, as normal, localised in momentum space. However, real probes are localised in both momentum and position. Probes can be narrower than background beams, and entirely miss them if not properly aligned. Neither of these situations can be described if the probe is taken to be a momentum eigenstate. Hence, localisation in position space becomes relevant, and we require a wavepacket for the probe. The full scattering amplitude to calculate is then
\be\label{SFI2}
	S_{fi} := \tint{p}\;\psi(p) \bra{l',\epsilon'}S\ket{p,\epsilon} \;,
\ee
in which $\ud{\sf p}$ is the Lorentz-invariant measure over the positive energy mass shell and the wavepacket $\psi$ obeys
\be
	\psi(p) = \Lambda(p)e^{ipx_0} \quad \text{with}\quad	\int\!\ud{\sf p}\, |\Lambda(p)|^2=1 \;. 
\ee
Here $x_0$ is the initial position about which the wavepacket is centred and $\Lambda(p)$ is sharply peaked (to be made precise below) around momentum $p_\mu = l_\mu$. The measure depends on three momentum coordinates, which can be ordinary vector momentum, but since we are dealing with photons it seems natural to take ${\sf p} = \{np,p^\perp\}$ defined with respect to some lightlike direction $n_\mu$. This corresponds to a foliation of spacetime into a time $nx$ and three spatial directions ${\sf x}$; this will be of use below. 

We first Fourier transform the background,
\be\label{Fourier}
	f_{\mu\nu}(x)=\int\!\ud^4k\; e^{-ikx}f_{\mu\nu}(k) \;,
\ee
and perform the three ${\sf x}$-integrals in (\ref{SFI2}), giving three delta-functions.  Because the external momenta are on-shell (three degrees of freedom), this is enough to determine a relation between the incoming and outgoing momenta; writing $\kappa = k +k'$, the sum of momenta coming from the two factors of $f_{\mu\nu}$, we find
\be\label{p-momentum}
	p=l'-\kappa-\frac{(l'-\kappa)^2}{2n(l'-\kappa)}n \;.
\ee
For the HIBEF experiment, the typical background momentum ($|{\bf k}|$, optical) is much lower than the electron rest mass, and also much smaller than the typical probe momentum ($|{\bf l}|$, x-ray), so $|{\bf k}| \ll |{\bf l}|$. We want to evaluate $S_{fi}$ under these assumptions.  We therefore make a low energy approximation typical when considering e.g.\ infra-red effects~\cite{Yennie:1961ad,Dittrich:1973rn}. In the exponent, we neglect quadratic (and higher) powers of the background momenta~\cite{IR}. Outside the exponent, we also expand to linear order in these momenta. Dropping $\{k,k'\}$ in the trace terms of (\ref{SFI}) corresponds to neglecting derivative terms which have in any case been neglected in deriving (\ref{LHE}); the result is that $F^\text{in}_{\mu\nu}\to l'_{[\mu}\epsilon_{\nu]}$ and $F^\text{out}_{\mu\nu}\to l'_{[\mu}\epsilon'_{\nu]}$. With this, the traces simplify considerably and we recover the structure in (\ref{worldline}). Finally, to be able to neglect $\{k,k'\}$ in the wavepacket $\Lambda$ we have to assume that
\be\label{L-condition}
	|(p-l')\partial_{l'}\Lambda(l')|\ll|\Lambda(l')| \;.
\ee
Since $p-l'=\mathcal{O}(k)$, (\ref{L-condition}) implies that the wavepacket cannot be too sharply peaked; its momentum space width $\Delta$ should be larger than the typical background momentum, $\Delta\gg |{\bf k}|$. What this means physically is that, in position space, the probe is localised at scales on which the background varies. This is seen explicitly by noting that, in the low energy approximation, (\ref{p-momentum}) becomes
\be\label{p-momentum-2}
	p = l' - \kappa + \frac{l' \kappa}{nl'}n \;,
\ee
so that when we undo the Fourier transformations (\ref{Fourier}), both $f_{\mu\nu}$ and the scattering amplitude become supported on a classical photon trajectory $x_{l'}^\mu$ as in (\ref{bana}):
\be
	\int\!\ud^4k f(k)\exp-ik\bigg(x_0+\frac{l'}{nl'}n(x-x_0)\bigg) =f(x_{l'}) \;.
\ee
A trivial reparameterisation trades $x_0$ for $x(0)$. Our assumptions have lead to each of the modes in the wavepacket being scattered forward~\cite{PW}. The amplitude becomes $S_{fi} =~\Lambda(l') \ddelta(x_{l'})$, which is the worldline integral~(\ref{worldline}), with path $x^\mu_{l'}$. The total probability of scattering with a polarisation flip is then
\be\label{P1}
	\mathbb{P}_\text{flip}=\tint{l}' |S_{fi}|^2 = \tint{l}'|\Lambda(l')|^2 |\ddelta(x_{l'})|^2 \;.
\ee
If we further assume, as normal, that the width of the wavepacket is small compared to the typical probe momentum ($\Delta\ll |{\bf l}|$) then we can as usual drop the wavepacket and integral from (\ref{P1}) and replace $l'_\mu \to l_\mu$, upon which the probability becomes
\be\label{final}
	\mathbb{P}_\text{flip}= \mathbb{P}_\text{forward+flip} =  |\ddelta(x_l)|^2\;.
\ee
\subsection{Quantum reflection}
To arrive at (\ref{final}) we assumed a separation of scales, namely that the characteristic frequency of the background is much smaller than that of the probe. It is interesting to ask what happens when this is not the case, and to compare with the quantum reflection calculation in~\cite{Gies:2013yxa}. Let the background now depend on a single spatial coordinate $x^1\equiv x$. Three of the integrals in (\ref{SFI}) can then be performed, giving a delta function supported on vector ${\bf p}= \{\pm l'_1 ,l'_2,l'_3\}$, describing forward ($+$) or back ($-$) scattering. Assuming the background polarisation is $x$-independent, the probabilities for forward scattering and reflection become, schematically,
\be\label{reflection}
	\mathbb{P}_\text{for.}\propto \bigg|\!\int\!\ud x\; f^2(x)\bigg|^2\;, \quad \mathbb{P}_\text{ref.}\propto \bigg|\!\int\!\ud x\; e^{2il_1x}f^2(x)\bigg|^2\;,
\ee
where the (different) proportionality constants depend on the probe momentum and polarisation, and the vector structure of the background. The reflection probability has the same structure as the reflection coefficient in~\cite{Gies:2013yxa}, and will be much smaller than the forward scattering probability {\it unless} the background has support for momentum on the order of the probe momentum. Hence, for proposed `optical + xray' laser experiments we expect photon reflection to be a small effect compared to birefringence. However, for other setups, as described in detail in~\cite{Gies:2013yxa}, it would be easier to look for the reflection signal, which has the advantage of being more easily separated from experimental noise.

\section{Examples in Gaussian beams and pulses}\label{SECT:BEAMS}
In the following sections we evaluate the flip amplitude for photons in various collision geometries with backgrounds modelling intense laser fields. (For reviews of classical and quantum physics in intense lasers see~\cite{EKK,DiPiazza:2011tq}.) In the context of vacuum birefringence, the amplitude $\ddelta$ is equal to the birefringence-induced ellipticity $\delta$, for probes which are sufficiently narrow compared to targets, see Sect.~\ref{SECT:DELTA}.

To evaluate the integral (\ref{worldline}) or (\ref{worldline2}) in a given $f_{\mu\nu}$, pick a momentum $l_\mu$, and path $x_l^\mu$ for the photon. Parameterise the path with $nx$, such that $n^2=0$ and $nl\not=0$. The two orthogonal photon polarisation vectors can be taken in any gauge. The line integral can then be calculated. To proceed, we need a pulse model.

The most common description of focused laser fields is a Gaussian beam in the paraxial approximation. Following~\cite{Davis:1979zz,McDonald:1995}, the paraxial beam can be defined by a wavelength $\lambda$ and focal waist $w_0$. These give the Rayleigh range $z_0=\pi w_0^2/\lambda$, and the beam divergence $\theta_0$ which we express as $s:=\tan\theta_0 = w_0/z_0$. Defining $\zeta : =1/(1+iz/z_0)$, the only nonzero field components are $B^y = E^x$, where
\be\label{E-parax}
	E^x_\text{parax} = \text{Re } E_0\,  e^{-i \omega(t-z)} \zeta e^{-\zeta \frac{r^2}{w_0^2}} \;,
\ee
and $E_0$ is the peak field strength, related to the cycle-averaged power by $P=\frac{\pi}{4}E_0^2 w_0^2$. The beam solves Maxwell's equations up to terms of $\mathcal{O}(s)$, as is made more explicit by measuring transverse position in units of $w_0$, writing $\rho := r/w_0$, and both longitudinal position and time in units of $z_0$, writing $z=z_0 \hat{z}$, $t=z_0\hat{t}$. Then
\be\label{E-parax-2}
	E^x_\text{parax} = \text{Re } E_0\,  e^{-i \frac{2}{s^2}(\hat{t}-\hat{z})} \zeta e^{-\zeta\rho^2} \;.
\ee
The first exponential is rapidly oscillating since $s\ll~1$. The second exponential is independent of $s$ and is slowly varying in comparison. The terms neglected in the paraxial approximation are~$\mathcal{O}(s)$.

Though the paraxial beam is easily understood, it is an unsatisfactory model. First, because it cannot describe a pulse: at any given point in space the field oscillates in time forever, without losing amplitude. Second, the energy in the beam is infinite~\cite{ENERGI}. The periodicity may not appear to be an issue, because our probe travels at $c$ and quickly passes into spatial volumes where the field is damped. However, as we will show explicitly, it is in fact essential to account for pulse duration. The simplest way to do so is to add to (\ref{E-parax}) a Gaussian envelope in $t-z$ (it is not enough to add an envelope in $t$) as so: 
\be\label{E-pulse}
	E^x_\text{pulse} = \text{Re } E_0\,  e^{-\tfrac{\Delta\omega^2}{4}(t-z)^2}e^{-i \omega(t-z)} \zeta e^{-\zeta \frac{r^2}{w_0^2}} \;,
\ee
in which $\Delta\omega$ is a frequency spread related to the FWHM duration of the pulse, $\tau_L$, by $\tau_L = {\sqrt{8\log 2}}/{\Delta\omega}$.  We require $\Delta\omega^2/\omega^2\ll 1$ for the field to be an approximate solution of Maxwell's equations. The first advantage of this `paraxial pulse' over the the paraxial beam (\ref{E-parax}) is that it is genuinely pulsed; the field is damped in all spacetime directions. The second advantage is that the pulse energy $\mathcal{E}$,
\be\label{energi1}
	\mathcal{E}  = \frac{1}{2}\int\!\ud^3{\bf x}\ ({\bf E}^2 + {\bf B}^2) \;,
\ee
is finite. Plugging (\ref{E-pulse}) into this expression and integrating out $r$ leaves, changing variable $\hat{z} \to u= (\hat{z}-\hat{t})/s^2$,
\be\begin{split}\label{SVP-MIG}
	\mathcal{E}  &= \frac{\pi E_0^2w_0^2}{2\omega}\!\int\!\ud u\ e^{-2\frac{\Delta\omega^2}{\omega^2}u^2}  \bigg[1 + \frac{\mathrm{C} + (s^2u+\hat{t})\mathrm{S}}{1+(s^2u+\hat{t})^2}\bigg] \;,
\end{split}
\ee
in which $\mathrm{C} = \cos 4u$ and $\mathrm{S} = \sin 4u$. The trig terms will be rapidly oscillating compared to the Gaussian (i.e.\ the envelope will belong to the slowly varying part of the field) since the spectral width obeys $\Delta\omega^2/\omega^2\ll 1$. We can then apply a slowly varying phase (SVP) approximation to the integral (\ref{SVP-MIG}), killing the trig functions and with them the time-dependent terms, leaving
\be\begin{split}
	\mathcal{E}  &\simeq \frac{\pi^{3/2}}{\sqrt{8}} \frac{E_0^2 w_0^2}{\Delta\omega} \;.
\end{split}
\ee
This is (within our approximation) constant, as the energy should be in a solution of Maxwell's equations. The limit $\Delta\omega\to 0$ ($\tau_L\to\infty$) recovers the infinite energy of the paraxial beam.  (The same result could be obtained starting with the energy density in (\ref{energi1}) and applying the SVP to a cycle-average over time $t$ before computing the integrals. While averaging is somewhat natural in periodic fields, the SVP can be applied more generally.)

The paraxial pulse (\ref{E-pulse}) can, unlike the beam (\ref{E-parax}), consistently account for the parameters of the proposed HIBEF vacuum birefringence bexperiment, see Table~\ref{TAB:PARAMS}~\cite{HIBEF,HP}. Expressing the energy in terms of power $P$ as	$\mathcal{E} =~\tau_L  P\sqrt{\pi /\log 16}$, and taking power, frequency and waist from Table~\ref{TAB:PARAMS} identifies $E_0~\simeq~2.99\times 10^{-4} E_S$. Given that the total energy is $30\,$J, we then find the FWHM pulse duration to be $\tau_L=~28.18\text{ fs}$, essentially the expected value. These parameters are used in the following calculations. Note that $\Delta\omega/\omega\simeq 0.04$, justifying the use of the SVP. The intensity distributions of the paraxial pulse and beam are shown in Fig.~\ref{FIG:PULSAR}.

The model (\ref{E-pulse}) is not an exact solution to Maxwell's equations. We show though in Sect.~\ref{VD} that all our results hold for more sophisticated models which are exact solutions. Thus the fine details of the model (e.g.\ higher orders in $s$) do not impact on our results. We therefore use here the simple model (\ref{E-pulse}), both for intuition and in order to provide some analytic results.  

We will now consider the effect on the amplitude of transverse impact parameter, incidence angle, timing jitter and probe beam shape. For an analysis of the role these variables play in elastic and inelastic photon-photon scattering in the collision of two intense pulses, see~\cite{King:2012aw}. (We emphasise that it is meaningful to talk about the amplitude here because, due to our polarisation choices, $T$ is real and related to the flip probability $\mathbb{P}$ via $T=\sqrt{\mathbb{P}}$.)

\begin{table}[t!]
\caption{\label{TAB:PARAMS} Optical laser parameters proposed for the HIBEF vacuum birefringence experiment~\cite{HP}.}
\begin{ruledtabular}
\begin{tabular}{l | r || l | r}
Wavelength $\lambda $& $800$~nm & Frequency $\omega$ & 1.55~eV\\ \hline  
Waist $w_0$ & $1.75~\mu$m  & Total energy & 30 J \\ \hline
Rayleigh $z_0$& 12 $\mu$m & Power & 1 PW \\ \hline
$s=w_0 / z_0$ & 0.15 & FWHM duration $\tau_L$ & 30 fs 
\end{tabular}
\end{ruledtabular}
\end{table}

\subsection{Transverse impact parameter}
In a birefringence experiment, probe and target beams would ideally be aligned so that their focal spots overlap. Here we illustrate the effect of impact parameters by considering a probe photon which reaches the focal plane of the Gaussian, $z=0$ , at the instant of peak field strength, $t=0$, but misses the focal spot (centred at the origin) by a transverse distance; this is the impact parameter $r$. Given the intensity distribution of our fields, we might expect that $\ddelta$ will fall as a Gaussian $\exp -2(r/w_0)^2$.

\begin{figure}[t!!!]
\centering\includegraphics[height=3.5cm]{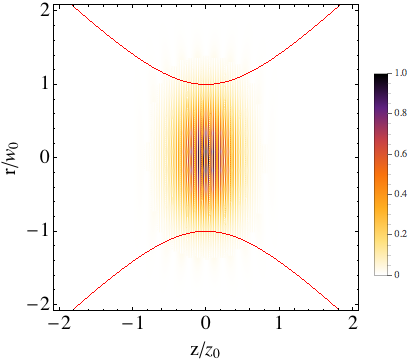}
\includegraphics[height=3.5cm]{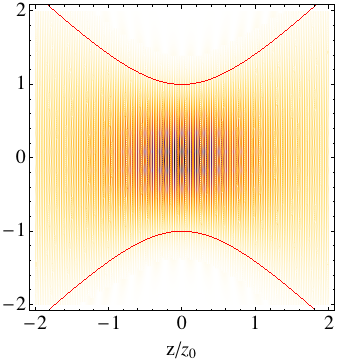}
\caption{\label{FIG:PULSAR} Intensity distributions at $t=0$, normalised to peak intensity. Left: the paraxial pulse~(\ref{E-pulse}). Right: the paraxial beam~(\ref{E-parax}), for which the intensity is periodic in time.}
\end{figure}

\begin{figure}[t!]
	\includegraphics[width=\columnwidth]{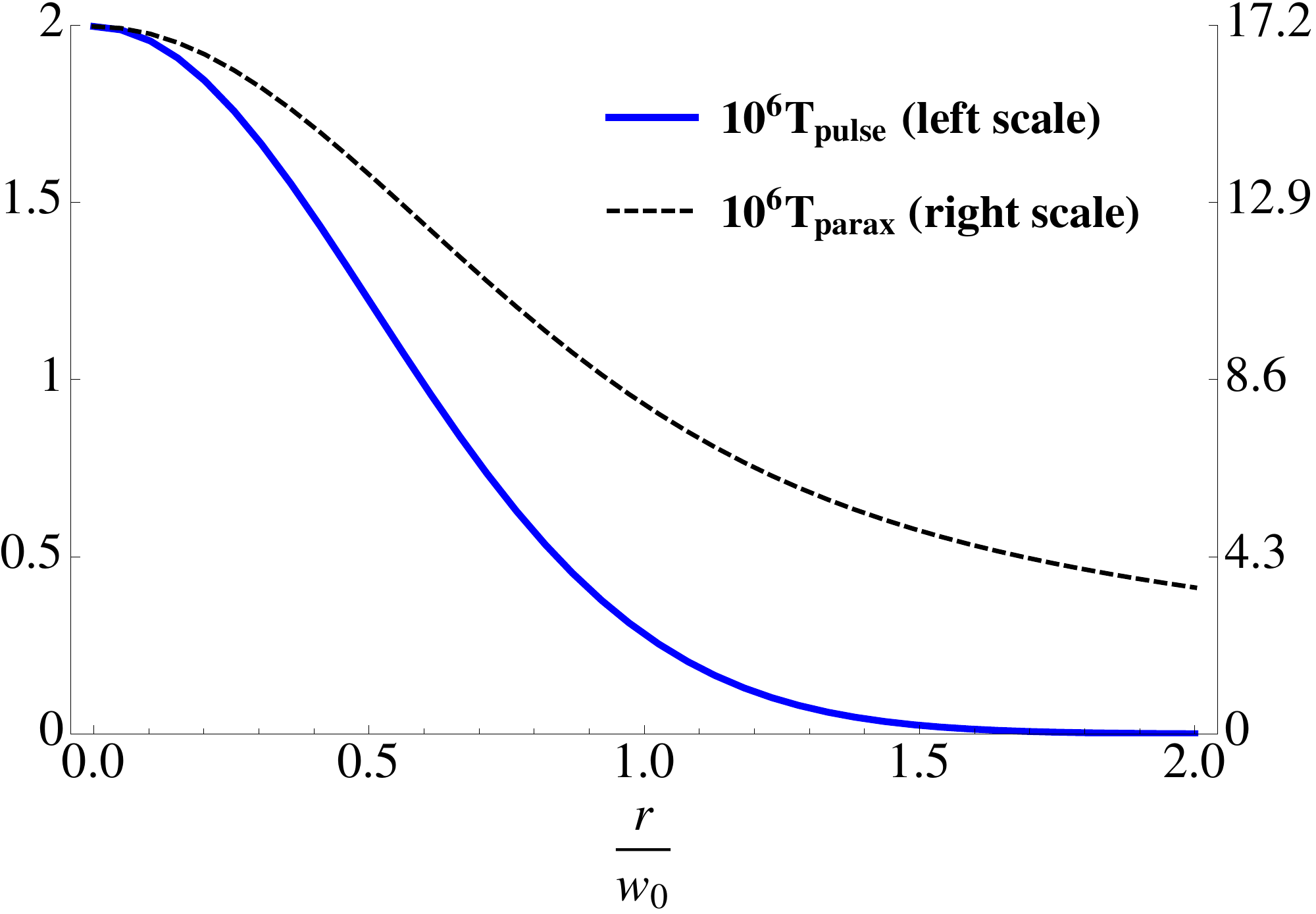}
	\caption{\label{FIG:R-I-PULSE} The scattering amplitude $\ddelta$ as a function of transverse impact parameter, $r/w_0 = \rho$. Parameters as in Table~\ref{TAB:PARAMS}. Note the different scales. The paraxial beam gives an order-of-magnitude overestimate and an unphysical enhancement at large $r$. (The approximations (\ref{delta-rho-approx}) and (\ref{short}) are indistinguishable from the exact results on the scale shown.)}
\end{figure}

We let the photon travel down the $z$-axis, so that $l^\mu = \omega'(1,0,0,-1)$. From here on, $\omega'=12.4$~keV assuming a hard X-ray photon~\cite{HIBEF,HP}. We parameterise with $\phi=nx$, $n^\mu=(1,0,0,1)$. The path is (with $\varphi$ the angle in the transvere plane)
\be\begin{split}\label{path1}
	x^\mu(\phi) &= \big\{\phi/2,r\cos\varphi,r\sin\varphi,-\phi/2\big\}  \;.
\end{split}
\ee
Taking a $45^\circ$ angle between the background and probe polarisations kills the second term in (\ref{worldline2})~\cite{Heinzl:2006xc}. The resulting line integral is easily performed numerically and the results are plotted in Fig.~\ref{FIG:R-I-PULSE}. Some analytic expressions are available to aid interpretation. We begin with the paraxial beam. Applying the SVP at to the worldline integral gives the following accurate approximation for the scattering amplitude $\ddelta$,
\be\label{delta-rho-approx}
	\ddelta(\rho) \overset{\text{parax}}{\simeq} \frac{\alpha}{15}
	\frac{E_0^2}{E_S^2} \frac{1}{s^2} \frac{\omega'}{\omega} 
	e^{-\rho^2} I_0 (\rho^2 ) \; , 
\ee
with $\rho := r/w_0$ and $I_0$ the  standard modified Bessel function. The resulting curve is indistinguishable from the numerically exact dashed curve in Fig.~\ref{FIG:R-I-PULSE} at the scale shown. The amplitude falls off if the impact factor is greater than the beam waist, $r>w_0$. This is natural given the spatial limits of the intensity distribution, see~Fig.~\ref{FIG:PULSAR}, but the falloff is much slower than might be expected; the Bessel function precisely cancels the exponential decrease leaving only a power law tail $ \sim 1/\rho$,
\be \label{asympt}
  \ddelta(\rho) \overset{\rho \gg 1}{\to} \frac{\alpha}{15} 
  \frac{E_0^2}{E_S^2}\frac{1}{s^2}  \frac{\omega'}{\omega}	 
  \frac{1}{\sqrt{2\pi} \rho} \;.
\ee
This would be a positive result, as such peripheral contributions could enhance e.g.\ birefringence signals. Unfortunately, it is unphysical, as we now show.

Assuming $\nu_0^2 := (s^2\omega/4\Delta\omega)^2 \ll 1$ (as holds for the HIBEF parameters where $\nu_0^2 \simeq 0.02$) the scattering amplitude $\ddelta(\rho)$ in a pulse is approximately given by
\be\label{short}
	\ddelta(\rho) \overset{\text{pulse}}{\simeq} \frac{\alpha}{15}
	\frac{1}{E_S^2}
	\frac{\mathcal{E}\omega'}{\pi^2 w_0^2} e^{-2\rho^2} \;.
\ee
To see what formula (\ref{short}) implies, consider Fig.~\ref{FIG:R-I-PULSE}.  We see that, for the same parameters aside from pulse duration, the paraxial beam overestimates the amplitude by almost an order of magnitude, for near head-on collisions. The reason is that, at fixed power, the paraxial beam is the `long pulse limit' of the pulse, $\tau_L\to\infty$ ($\Delta\omega\to 0$) {\it only} under the assumption that the pulse energy is allowed to increase to infinity. This is unphysical, but is what is implicitly assumed when using paraxial beams.

Even if one tries to compensate by artificially reducing the field strength, we see directly from Fig.~\ref{FIG:R-I-PULSE} that the behaviours of the amplitudes are still very different; (\ref{short}), in contrast to (\ref{asympt}), does have an exponential tail, with the same Gaussian fall-off as the intensity distribution. If we rewrite (\ref{short}) in terms of peak field strength,
\be\label{short2}
	\ddelta(\rho) \overset{\text{pulse}}{\simeq} 
	\frac{\alpha}{15} \frac{E_0^2}{E_S^2} \frac{1}{\sqrt{8\pi}}
	 \frac{\omega'}{\Delta \omega} e^{-2\rho^2}
	\;,
\ee
then it is easy to compare the large impact parameter behaviour of $T(\rho)$ in the paraxial beam and pulse. Asymptotically one finds
\be
	\frac{T_\text{pulse}}{T_\text{parax}} \overset{\rho\gg 1}{\rightarrow} 2\nu_0 \rho \exp(-2\rho^2) \;.
\ee
At large impact parameter $\rho$, the amplitude in a pulse is exponentially suppressed compared to that in a paraxial beam, and the `enhanced signal' seen above is lost.

More physically one can imagine, at fixed energy, compressing/stretching the pulse to increase/reduce the peak amplitude. Provided the pulse remains short, and (\ref{short}) applies, such variations give a minimal effect, since we see from (\ref{short}) that $\ddelta(\rho)\sim \mathcal{E}$, fixed. In such a situation, and as predicted in~\cite{Dinu:2013gaa}, it is the total energy of the pulse which is relevant to helicity flip and birefringence.

\subsection{Angle of incidence}
\begin{figure}[t!]
	\includegraphics[width=\columnwidth]{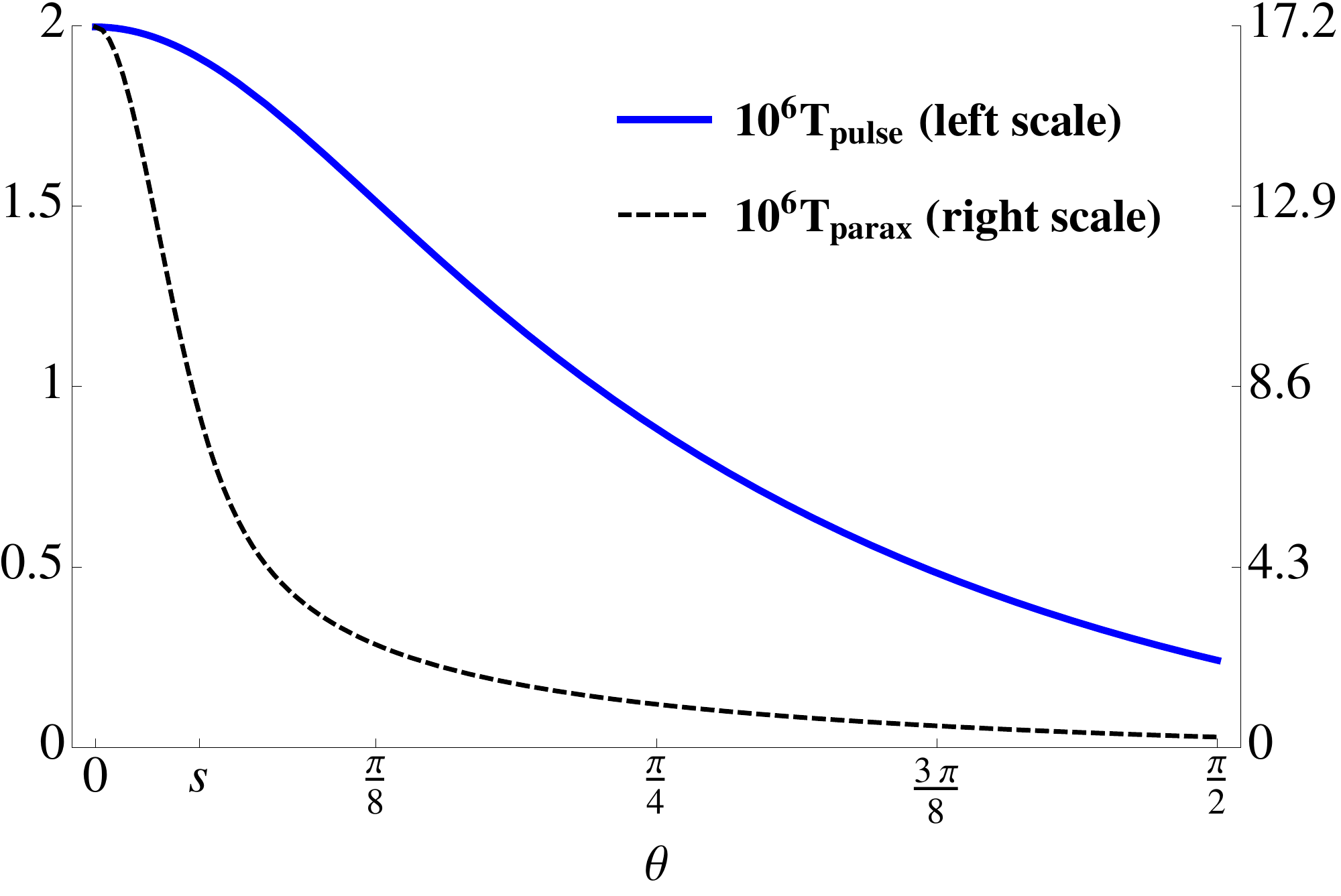}
	\caption{\label{FIG:VINKEL-I-PULSE} The scattering amplitude $\ddelta$ as a function of incidence angle $\theta$. Parameters as in Table~\ref{TAB:PARAMS}. Note the different scales. The paraxial beam model (black/dashed) overestimates the signal. However, $\ddelta$ is much less sensitive to incidence angle in a pulse (solid/blue) than in the paraxial beam.}
\end{figure}

%
As a second example, consider a collision with an acute incidence angle $\theta$ between the probe and beam axes (where $\theta=0$ is head-on). We again take the best case scenario regarding polarisations, such that the second term in (\ref{worldline2}) vanishes. The results are plotted in Fig.~\ref{FIG:VINKEL-I-PULSE}. We see immediately that, unlike the case of impact parameter, the amplitude is much {\it less} sensitive to collision angle once pulse duration is accounted for. In the paraxial beam, the amplitude drops quickly when the collision angle exceeds the beam divergence, $\theta > \tan^{-1}s\simeq s$, which again is natural.  (As a function of $s^{-1}\tan\theta\sim\theta/s$ rather than $r/w_0$, the curve is almost identical to that for impact parameter in the beam case, Fig,~\ref{FIG:R-I-PULSE}.)  In the pulse, though, the signal drops much slower, extending all the way to transverse collision angle $\theta=\pi/2$; indeed, see~\cite{DiPiazza:2006pr} for proposals to measure induced probe ellipticity and polarisation rotation based on transverse collisions.

The reason for the reduced sensitivity to angular incidence is as follows. While the paraxial beam is effectively a time-independent distribution, vanishing outside a spatial region, the pulse effectively exists only at the origin for a brief instant, and is otherwise gone. This means that, provided the probe arrives at the origin at the right instant in time, the angle of incidence is relatively unimportant and a small deviation from a head-on collision will not significantly reduce the amplitude.  This insensitivity to incidence angle is a positive result, as it indicates a certain robustness of the amplitude. It does though raise the question of what happens when the pulse arrives early or late to the focal spot, missing the instant of peak field strength; this is considered in the next subsection.

The paraxial beam model overestimates the pulse result, though the degree of overestimation is angle--dependent.  This is relevant in the light of experimental proposals based on transverse collisions, and in the interests of simplifying calculations. We begin with head-on collisions, in which the probe sees the longitudinal extent of the beam. We have used above (\ref{short}) that $\nu_0^2\ll 1$ which implies $\tau_L/z_0\ll \sqrt{32\log 2}\simeq 4.7$. Hence the pulse duration must be much less than the Rayleigh range, and it is clear that the paraxial beam will not give an accurate description of the physics. If we define $\Upsilon(\theta) = \ddelta(\theta)_\text{parax}/ \ddelta(\theta)_\text{pulse}$ then an analytic estimate for the degree of overestimation is easily found,
\be
	\Upsilon(0) \simeq 4\sqrt{\pi \log2}\,\frac{z_0}{\tau_L} \;.
\ee
This gives $\Upsilon(0)\simeq 8.6$ for HIBEF parameters, an almost order of magnitude overestimate in agreement with Fig.~\ref{FIG:VINKEL-I-PULSE}. For transverse collisions, $\theta=\pi/2$, we find
\be
	\Upsilon(\pi/2) \simeq \sqrt{1 +\frac{w_0^2}{\tau_L^2} \log4} \;,
\ee
implying that the pulse duration must be greater than the beam width in order for the paraxial beam model to be accurate. For  Fig.~\ref{FIG:VINKEL-I-PULSE}, $\Upsilon(\pi/2)\simeq 1.0$, so that there is almost no overestimation. However, the actual amplitude at transverse collision is reduced by a factor $8.3$ compared to the head-on scenario.

\subsection{Timing jitter and competing effects}
\begin{figure}[t!]
	\includegraphics[width=\columnwidth]{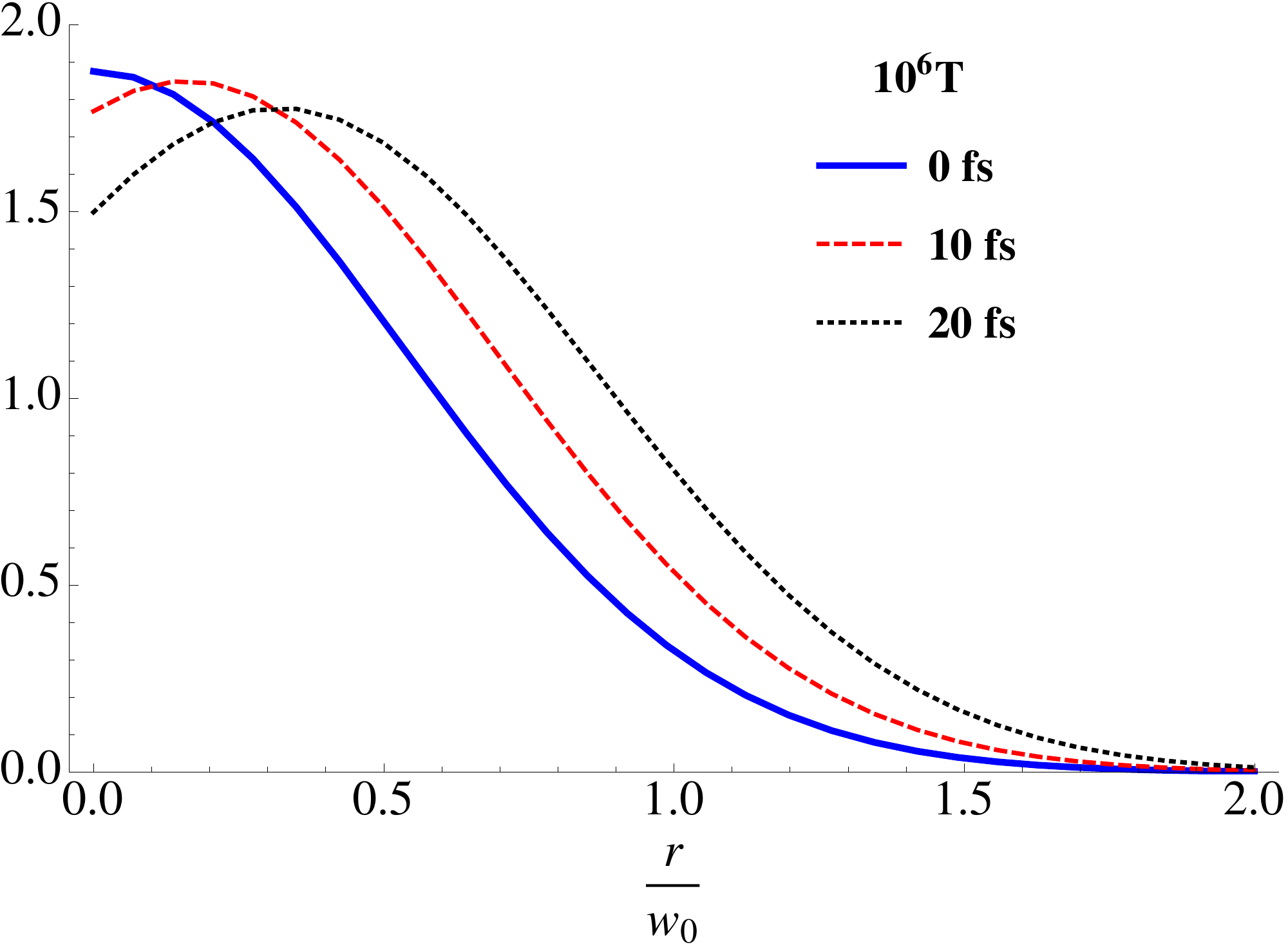}
	\caption{\label{FIG:ALL} The probe arrives at the focal plane $\{0,10,20\}\,\text{fs}$ after the field has peaked, with incidence angle $10^\circ$ and azimuthal angle $180^\circ$. Introducing a nonzero impact parameter then improves the signal.}
\end{figure}

Finally, we can (somewhat roughly) model the impact of `timing jitter' in e.g.\  triggering laser pulses, by considering a probe which misses the focal spot in time, as well as in space. Jitter alone naturally reduces the amplitude, as does a combination of jitter and nonzero impact parameter; if at $t=0$ the photon is not at $z=0$ but $z = 2z_0\tau$ and $r=w_0\rho$ then
\be\label{jitt}
	\ddelta(\rho,\tau)\simeq \frac{\ddelta(0,0)}{1+\tau^2} e^{-2\rho^2/(1+\tau^2)} \;.
\ee
If multiple sources of signal reduction are known to be present, though, introducing another can actually improve the signal. Assume for example that a collision angle of $10^\circ$ is required experimentally, and that a timing issue results in the probe arriving late to the focal spot. Under such conditions, deliberately introducing e.g.\ an impact parameter can increase the amplitude, as is shown by the dashed and dotted curves in Fig.~\ref{FIG:ALL}. If $\sigma$ is the angle between $\mathbf{x}^\LCperp_0$ and $\mathbf{l}^\LCperp$, then the approximate behaviour of the amplitude is given by (\ref{jitt}) with $\rho^2$ replaced by
\be
	\rho^2-2\rho \tau \theta/s \cos(\sigma) + (\tau \theta/s)^2 \;,
\ee
which, depending on parameter values and signs, can describe a shift of the Gaussian as seen in Fig.~\ref{FIG:ALL}.

\subsection{Ellipticity}\label{SECT:DELTA}
Ellipticity is the most commonly considered signal of vacuum birefringence. It is perhaps more natural though, from both scattering and experimental perspectives, to consider  the number of photons which could pass through a given arrangement of polarisers and be detected (a detailed calculation of which will be presented elsewhere). Hence we will discuss the ellipticity only briefly.
 
In~\cite{Dinu:2013gaa} we showed, for the case of plane wave probes (with a single frequency) that the induced probe ellipticity $\delta$ is equal to the amplitude $T$. The ellipticity for beam-like probes (with a frequency range) is given by averaging $T$ over the transverse distribution of photons in the probe beam. This can be shown explicitly by combining the methods developed in~\cite{Dinu:2013gaa}, in which the probe field is obtained from the expectation value of the field operator, with the Heisenberg-Euler approach of Sect.~\ref{SECT:HE}, or by solving the modified Maxwell equations following from the Heisenberg-Euler Lagrangian, as in \cite{DiPiazza:2006pr,King:2012aw,King:2013zz}. For the purposes of this work it is sufficient to illustrate the situation as follows. Consider the specific case of two paraxial pulses (as above) which are counterpropagating, up to a transverse separation $r$. One pulse is the optical target, the second is the probe. For a paraxial pulse probe with focal width $\omega'_0$ the transverse photon distribution is Gaussian, and the ellipticity becomes
\be\label{ellintT}
	\delta=\int\frac{\ud^2 y^\LCperp}{\pi {w_0'}^2}e^{-(y_\LCperp/w'_0)^2}T[x_\gamma]\;,
\ee
with the wordline given by, compare (\ref{path1}),
\be
	x^\mu_\gamma(\phi)=\{\phi/2,y^1 + r\cos\varphi,y^2+r\sin\varphi,-\phi/2\} \; , 
\ee
in which $y^\LCperp$ parameterises the probe width and $r$ is the transverse separation. If the probe waist $w'_0$ is small compared to the scales at which the background varies, we can simply neglect $y^\LCperp$ in $T$, and the Gaussian integrals in (\ref{ellintT}) can be performed, leaving $\delta = \ddelta$; the ellipticity is then equal to the flip amplitude again. For wider probes there will be corrections to $\delta$, given by (\ref{ellintT}). 

\begin{figure}[t!]
	\includegraphics[width=\columnwidth]{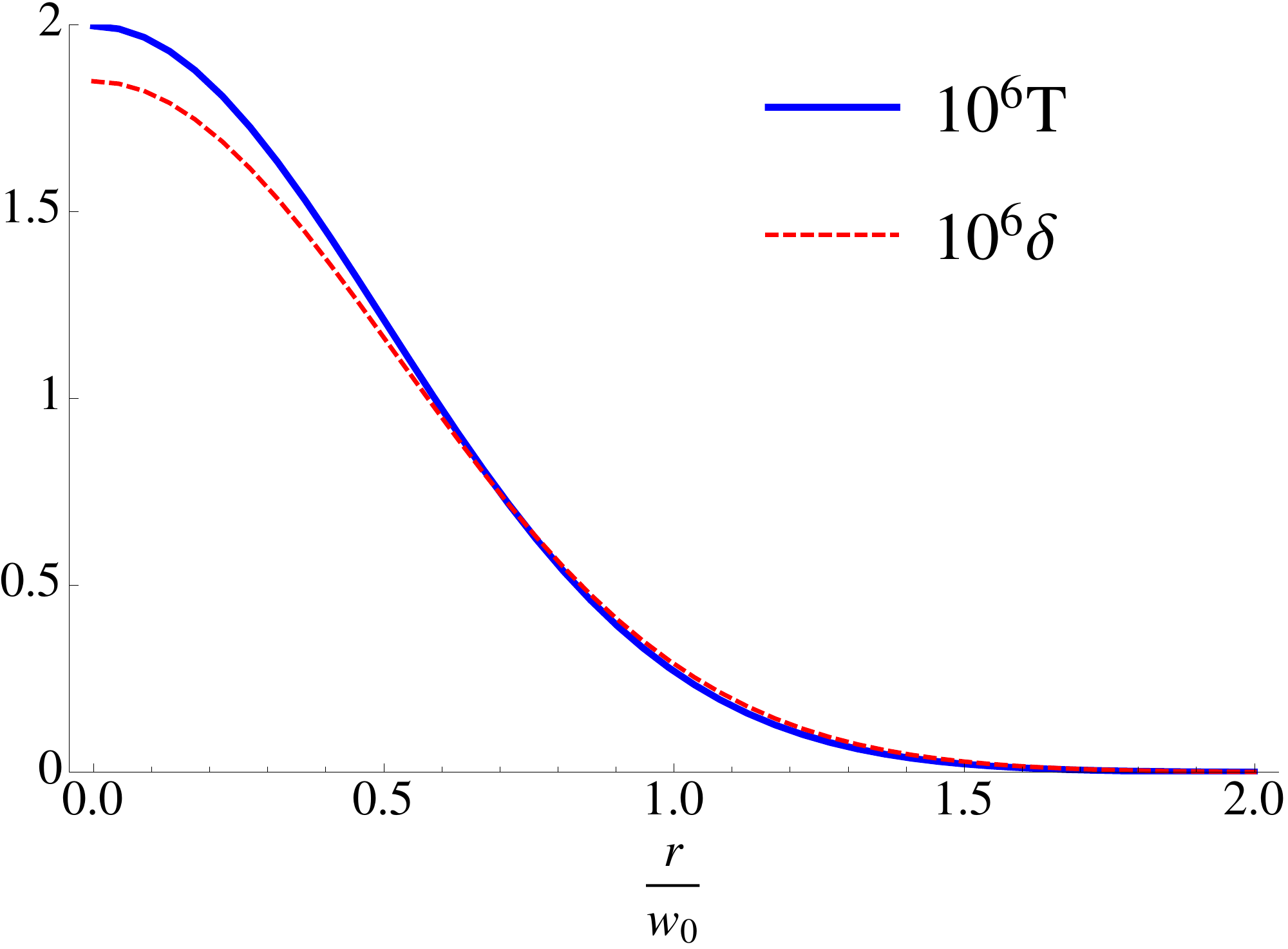}
	\caption{\label{FIG:DELTA} The amplitude and ellipticity for the HIBEF parameters, as a function of the transverse beam separation (in units of target waist radius) in an otherwise head-on collision.} 
\end{figure}

In particular, the expected HIBEF probe width is $w_0'=0.3 \; \mu$m, which is not so much less than the target waist of $w_0=1.75 \; \mu$m; we should therefore expect some finite width effects. Given the accuracy of (\ref{short}), we can use that approximation to calculate (\ref{ellintT}) analytically. Defining the ratio of probe to target widths $\varpi = 2{w'_0}^2 / w_0^2$ we find
\be\label{shortdelta}
	\delta(\rho) \simeq T(\rho) \frac{1}{1+\varpi} e^{2\rho^2\frac{\varpi}{1+\varpi}}\;.
\ee
For our parameters we have $\varpi \simeq 0.08$, so that $\delta(\rho)\simeq (1-\varpi) \ddelta(\rho) = 0.92 \, \ddelta(\rho)$ for small $\rho$, and there are indeed finite-size corrections to the narrow probe result, as shown in Fig.~\ref{FIG:DELTA}. Note that the full exponent in (\ref{shortdelta}) is~$-2\rho^2/(1+\varpi)$, so that the falloff of the ellipticity is still Gaussian. See~\cite{King:2013zz} for the ellipticity in a `double slit' setup, in which a probe passes through two parallel, intense optical fields; the probe ellipticity there also exhibits a Gaussian falloff as a function of the separation between the optical lasers.

\section{Exact solutions}\label{VD}

Finally, we confirm that the above results hold in more sophisticated pulse models which are, in particular, exact solutions of Maxwell's equations. We base our analysis on the Narozhny-Fofanov beam~\cite{NF} (see also~\cite{Fedotov}), describing the background optical laser as a momentum distribution peaked around $k^\mu=\omega(1,0,0,1)$, for propagation in the $z$-direction. To describe a pulse we take a distribution $\Psi(|{\bf k}|)$ in $|{\bf k}|\in\mathbb{R}^\LCp$, and to describe focussing we take a vectorial distribution ${\bf \Phi}(\mathbf{n})$ on the photons' direction $\mathbf{n} \in \mathbb{S}^2$. A gauge potential  (in radiation gauge, $A^0=0 = \partial^i A^i $) is
\be\label{modl}
	{\bf A}(x) =  \mathcal{A}\, \text{Re}\int\!\ud^3 k \ \Psi(\omega){\bf \Phi}(\mathbf{n})e^{-ikx} \;,
\ee
in which $\mathcal{A}$ is an amplitude, $k^2=0$ and ${\bf n}= {\bf k}/|{\bf k}| = \{\sin\theta \cos\phi, \sin\theta \sin\phi, \cos\theta\}$. In~\cite{NF}, $\Psi$ was chosen to be a delta function in $|{\bf k}|$, giving a single frequency component to the beam, and the angular distribution in ${\bf \Phi}$ was limited by a step function, i.e.\ a `hard cutoff'.

\begin{figure}[t!!]
	\includegraphics[width=\columnwidth]{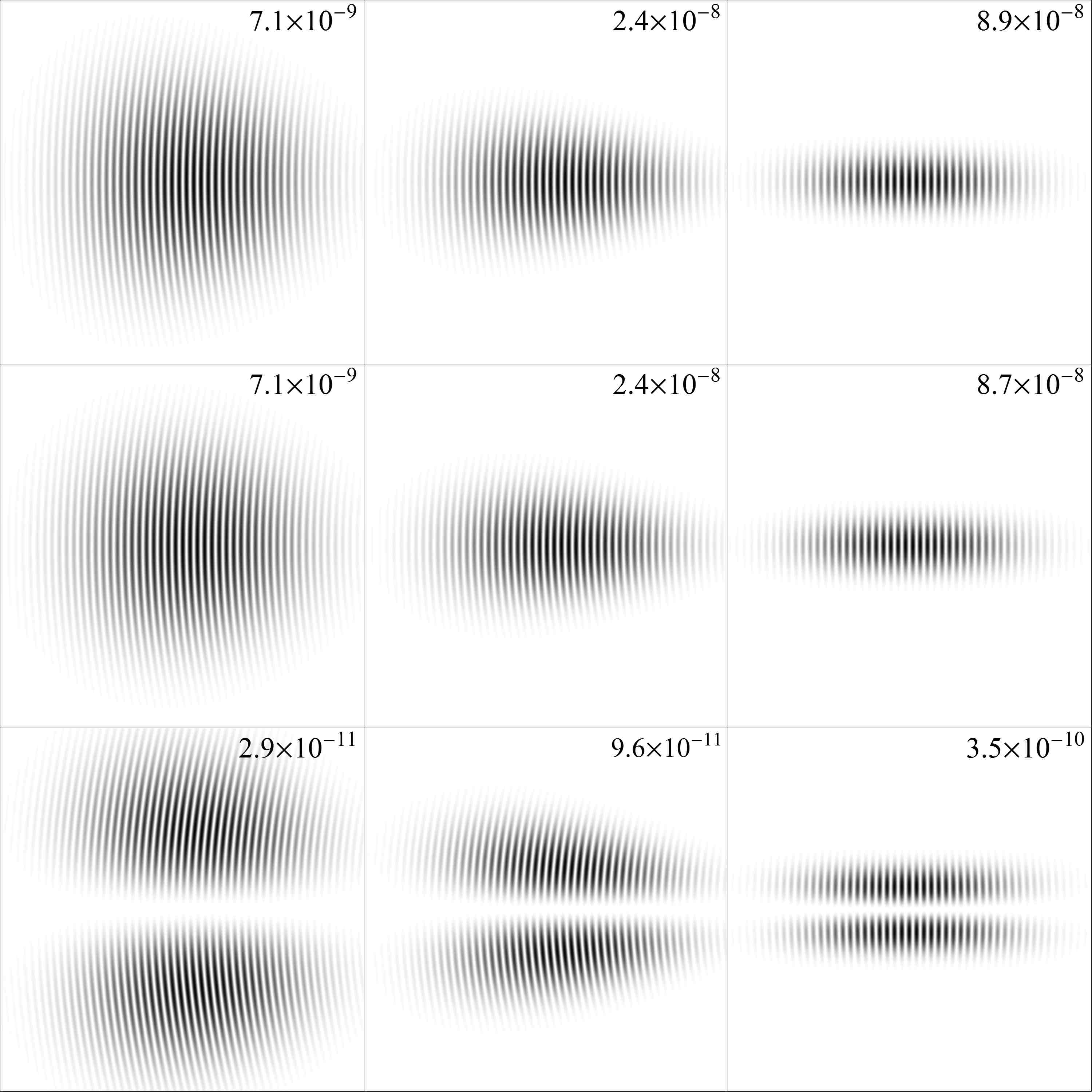}
	\caption{\label{FIG:VD-COMP} $E^2_x / E_S^2$ (peak values shown in each panel) for the Gaussian pulse (\ref{E-pulse}) (first row) and the exact solution (\ref{modl})-(\ref{modl3}) (second row). The third row shows longitudinal $E^2_z/E_S^2$ of the exact solution. The cross section is made through the $x$--$z$ plane, where $E_z$ is largest. The three snapshots are taken at $66$~fs/ $20\,\mu m$ apart, the rightmost being at the focus.}
\end{figure}

We choose a space fixed angular distribution, so that we can eliminate ${\bf E}_y$ from the outset, giving linear ($x$) polarisation of the electric field in the plane transverse to propagation. This is analogous to the Gaussian pulse above, and mirrors the function of a real polariser. The price we pay is that the electric field will develop a longitudinal component $\mathbf{E}_{z}$, unlike in~\cite{NF} where it can be taken purely transverse. When the angular spread is small, so is $\mathbf{E}_{z}$, matching what happens in a Gaussian beam when $\mathcal{O}(s)$ corrections are included. To avoid edge effects, we take bump-function distributions in frequency and angle which, for the focussing parameters we consider, are very close to Gaussian distributions. Explicitly, we have
\be\label{modl2}
	\Psi(|{\bf k}|) = \exp\bigg[ - \bigg(\frac{2 \omega}{\pi \sigma_\omega}\bigg)^2 \tan^2  \bigg(\frac{\pi(|{\bf k}| - \omega)}{2\omega}\bigg)\bigg]
\ee
with $|{\bf k}|\in[0,2\omega]$, and 
\be\label{modl3}
	{\bf \Phi}(\mathbf{n}) =  
	\exp \bigg(-\frac{\tan^2\theta}{\sigma^2_{\theta }}\bigg)\, 
	\hat{\mathbf{y}}\times \mathbf{n (\theta, \phi)}\;,
\ee
where $\hat{\mathbf{y}}$ denotes the unit vector in the $y$-direction. The larger $\sigma_{\theta}$, the more focused the beam and the smaller the focal waist. To parallel the discussion above, we choose $\sigma_\theta=s$ and $\sigma_{\omega}=\Delta\omega$. Taking the total energy to be $30$J then determines the amplitude~$\mathcal{A}$. 

\begin{figure}[t!!!]
	\includegraphics[width=0.9\columnwidth]{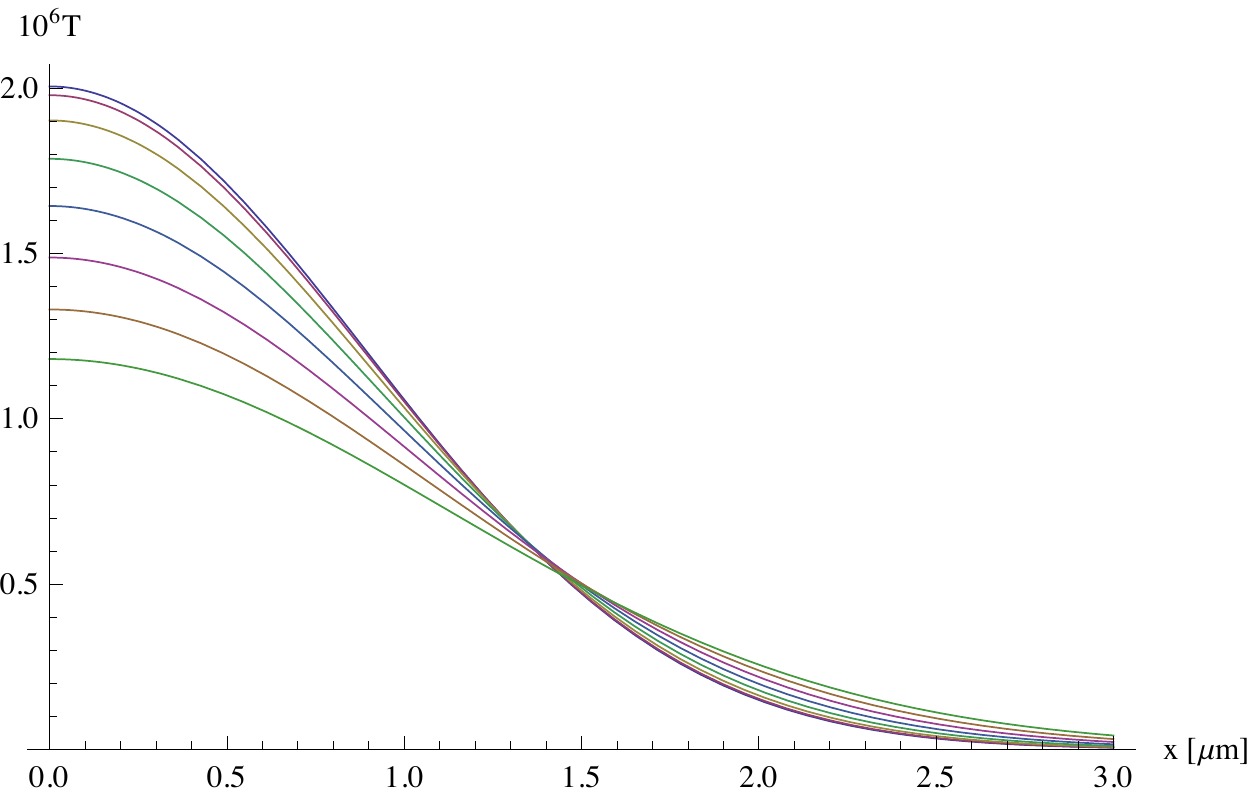}
	\caption{\label{FIG:VD-RHO} The scattering amplitude $\ddelta$ for a probe with momentum anti-parallel to $z$, which passes through the focal plane at time $t$, at a distance $x$ from the focus. Top to bottom, $t=\{0, 10, 20\ldots 70\}$ fs.}
\end{figure}

Though it is not possible to perform all the integrals in (\ref{modl}) analytically, this exact solution of Maxwell's equations looks very similar to (\ref{E-pulse}) in position space, as shown in Fig.~\ref{FIG:VD-COMP}. Far from the focus the wavefronts are circular, centred at the focus. The transverse field has cylindrical symmetry around $z$, while the small longitudinal field is proportional to $\cos\phi$. However, the $\phi-$dependent effects introduced are $\mathcal{O}(1\%)$, and we do not show them here.

\begin{figure}[t!]
	\includegraphics[width=0.9\columnwidth]{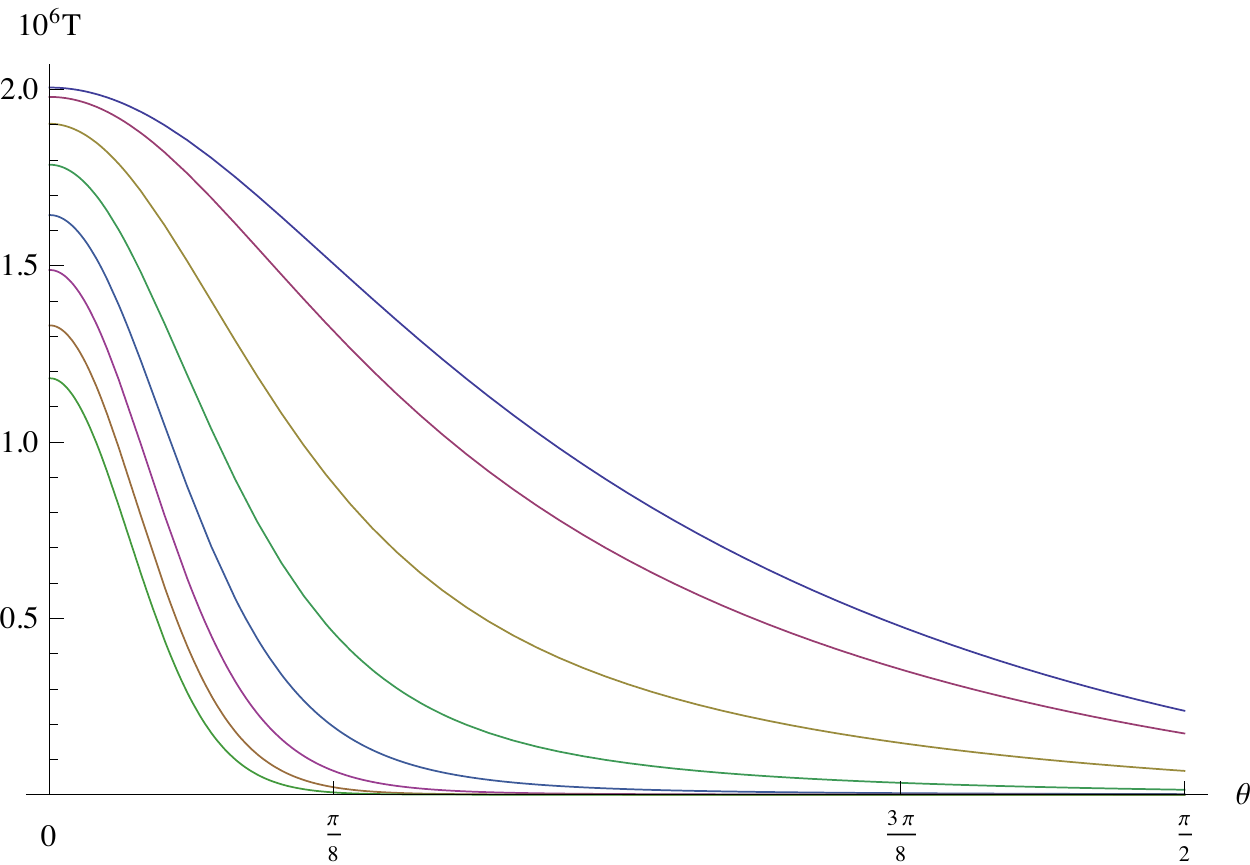}
	\caption{\label{FIG:VD-ANGLE} The scattering amplitude $\ddelta$ for a probe with incidence angle $\theta$ to the $z$-axis, which passes through the focus at (top to bottom) times $t=\{0, 10, 20\ldots 70\}$ fs.}
\end{figure}

\begin{figure}[t!!]
	\includegraphics[width=0.9\columnwidth]{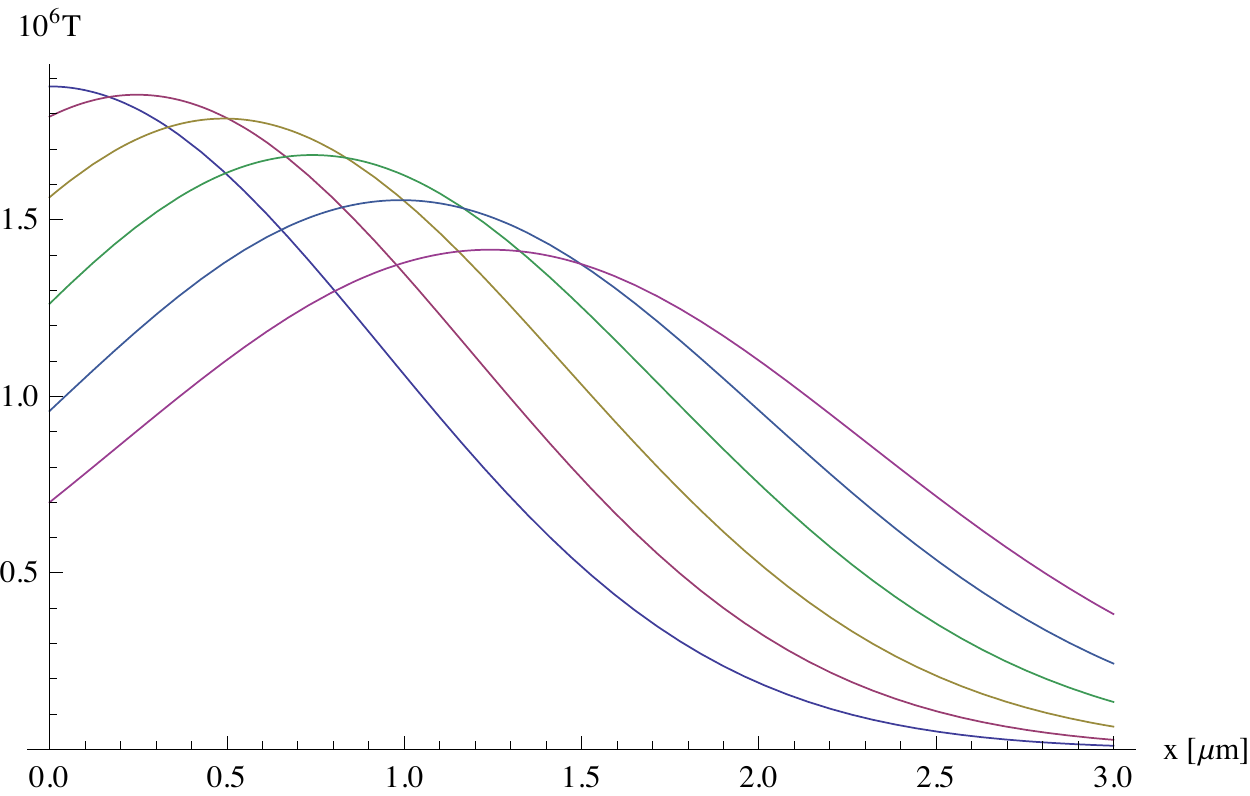}
	\caption{\label{FIG:VD-JITTER}  The probe arrives at the focal plane a distance $x$ from the focal point, at $\{0,10\ldots 50\}\,\text{fs}$ after the field has peaked, with incidence angle $10^\circ$ and azimuthal angle $180^\circ$. As in Fig.~\ref{FIG:ALL}, a nonzero impact parameter improves the signal.}
\end{figure}

In Fig.s~\ref{FIG:VD-RHO}--\ref{FIG:VD-JITTER} we plot $\ddelta$ as a function impact parameter, angle and jitter, with parameters as above. The results are practically identical to those obtained for Gaussian pulses, both in amplitude and form, implying that our results are insensitive to the fine details of the pulse model. 

\section{Conclusions}\label{SECT:CONCLUSIONS}

We have found a simple expression for the low energy photon helicity/polarisation flip probability in arbitrary background fields. The result can be deduced from that in a plane wave background, by observing that the lightfront time integral therein can be interpreted as an integral over the worldline of a massless particle. This is another example of how lightfront field theory is well suited to studying strong field QED~\cite{Neville:1971uc,Bakker:2013cea}. A derivation from Heisenberg-Euler, although more involved, gives insight into the approximations behind the result, and how the energy scales in play relate to the relevance of forward vs. back-scattering in laser-laser collisions.

The flip probability is closely related to the ellipticity to be measured in the proposed vacuum birefringence experiment at HIBEF~\cite{Heinzl:2006xc,HP}. Our results therefore give us a simple method for investigating the impact of beam geometry on birefringence signals. We have seen that beam models which do not account for pulse duration (such as standard paraxial Gaussian beams) overestimate both the flip amplitude (by an order of magnitude) and the relevance of peripheral collisions. In short pulses, the signal reduction due to `imperfect' collision angle is much less severe than predicted by the paraxial beam model, provided the probe is timed to arrive at the focus at close to the instant of peak field strength. While the effect of any single imperfection (collision angle, impact parameter, jitter) naturally reduces the signal, we have also seen that if it is experimentally necessary to include e.g.\ an angle, then it may be possible to optimise other parameters to partially counter its negative effect.

\subsection*{Acknowledgements}
The authors are supported by the strategic grant POSDRU/159/1.5/S/137750 (V.D.), the Swedish Research Council, contracts 2011-4221 (A.I. and G.T.), 2010-3727 and 2012-5644 (M.M.), and by the European Science Foundation framework {\it Super-Intense Laser-Matter Interactions}, grant 6481 (A.I.). A.~I. thanks V.~Florescu, M.~Boca and the Dept.~Physics, Bucharest-M\u agurele for hospitality. T.H.\ thanks H.P.\ Schlenvoigt for pointing out the relevance of timing jitter.

\appendix


\begin{thebibliography}{99}
\bibitem{ZMP}
This parameterisation cannot be used if $n_\mu \propto l_\mu$, for then $ln=0$; this is a manifestation of the zero mode problem of lightfront quantisation~\cite{Brodsky:1997de,Heinzl:2000ht}. In this case, the problem is avoided simply by choosing a different $n_\mu$.
\bibitem{PW}
We note in passing that for plane wave backgrounds one can take $n\propto k$, and (\ref{p-momentum}) reduces to $p=l'$. The condition (\ref{L-condition}) is then satisfied automatically and the photon always scatters forward.
\bibitem{ENERGI}
It is sometimes stated that paraxial beams have finite energy; this is a misnomer, referring only to a finite energy density obtained after integrating over transverse $r$.
\end{thebibliography}
\end{document}